\documentclass{statsoc}

\usepackage[left=1in,top=1in,right=1in,bottom=1.4in,head=1in]{geometry}

\usepackage{graphicx}
\usepackage{amsmath,amssymb}

\usepackage{etoolbox}

 \usepackage[utf8]{inputenc}

\usepackage[normalem]{ulem}

\usepackage{natbib}
\usepackage[english]{babel}

\usepackage[utf8]{inputenc}
\usepackage{multicol}
\usepackage{multirow}

\usepackage{hyperref} 
\usepackage{color}
\usepackage{colortbl}
\usepackage[space]{grffile}
\usepackage{bbm}
\usepackage{mathtools}
\RequirePackage{algorithm}
\RequirePackage{algorithmic}
\usepackage{bbm}

\newcommand{\E}{\mathbb{E}}

\newcommand{\Var}{\mathrm{Var}}
\newcommand{\Cov}{\mathrm{Cov}}

\newcommand{\zbold}{\boldsymbol{z}}
\newcommand{\hbold}{\boldsymbol{h}}
\newcommand{\sbold}{\boldsymbol{s}}
\newcommand{\ybold}{\boldsymbol{y}}
\newcommand{\thetabold}{\boldsymbol{\theta}}

\hypersetup{
    colorlinks,
    citecolor=black,
    filecolor=blue,
    linkcolor=blue,
    urlcolor=blue
}

\addto\captionsenglish{\renewcommand*\abstractname{Summary}}

\title[]{Resampling Methods for Detecting Anisotropic Correlation Structure}
\author[Author 1 {\it et al.}]{Assaf Rabinowicz\thanks{\scriptsize
This work was supported by the \textit{Israel Science Foundation, grant 1804/16} and by the \textit{European
Union Seventh Framework Programme
grant agreement no. 785907 (Human Brain Project)}}}
\address{Tel Aviv University, Tel Aviv, Israel.}
\email{assafrab@gmail.com}
\author[ ]{Saharon Rosset}
\address{Tel Aviv University, Tel Aviv, Israel.}
\email{saharon@tauex.tau.ac.il}

\begin{document}
\hspace{10cm}



\begin{abstract}
    This paper proposes parametric and non-parametric hypothesis testing algorithms for detecting anisotropy --- rotational variance of the covariance function in random fields. Both algorithms are based on resampling mechanisms, which enable avoiding relying on asymptotic assumptions, as is common in previous algorithms. The algorithms' performance is illustrated numerically in simulation experiments and on real datasets representing a variety of potential challenges.
\end{abstract}
\noindent%

\newpage

\section{Introduction}\label{section: introduction}
Estimating the covariance function has an essential role in various fields involving spatial data analysis, such as in climategoraphy, ecology and neurosceince \citep{hohn1998geostatistics,zhong2008classifying,mihoub2016modeling}. A common example for estimating the covariance function is in fitting Gaussian process regression (GPR) for predicting spatial data \citep{rasmussen2003gaussian}. However, estimating the covariance function is also required in many other settings, including unsupervised learning tools, such as in dimensional reduction using Gaussian process latent variable models \citep{li2016review}. 

Commonly, the covariance function of a sample $\zbold(S)=[z(\sbold_1),...,z(\sbold_n)]$ --- where $z(\sbold_i)\coloneqq z_i\in\mathbb{R}$ is sampled from the random field $\mathcal{Z}(\mathcal{S})$ at the coordinate values $\sbold_i\in\mathcal{S}\subset\mathbb{R}^{q}$ ---
is estimated using a \textit{kernel function},  $\mathcal{K}_{\thetabold}(\sbold_i-\sbold_j):\; \mathbb{R}^q\to \mathbb{R},$ where $\thetabold\in\Theta$ is the kernel's parameters. A popular kernel function is the exponential kernel:
\begin{align}\label{Kernel: exponential, isotropy}
\sigma_{s}^2\times\exp{\big(-\frac{\|\hbold_{i,j}\|}{\lambda}\big)+\sigma^2_{\epsilon}},    
\end{align}
where $\hbold_{i,j}=\sbold_i-\sbold_j,$
$\sigma_{s}^2\in \mathbb{R}^{+}$ is the signal parameter, $\lambda\in \mathbb{R}^{+}$ is the length-scale parameter and $\sigma^2_{\epsilon}\in \mathbb{R}^{+}$ is the variance of the independent error term. In this case, $\thetabold=[\text{exponential kernel},\sigma_{s}^2,\lambda,\sigma^2_{\epsilon}].$\footnote{Commonly, the variance of the error term, $\sigma^2_{\epsilon},$ which does not depend on the distance is not included in the kernel and therefore the covariance function is the kernel plus the error term, however in order to simplify writing the error term is included in the kernel function.}

The exponential kernel, as well as many other kernels, assumes weak stationarity, i.e.,
\begin{itemize}
    \item $\E z(\sbold)=\E z(\sbold+\sbold_{\tau})=\mu,$ where $\mu\in \mathbb{R}$ and $\sbold_{\tau}\in \mathbb{R}^{q},$ such that $\sbold+\sbold_{\tau}\in\mathcal{S}.$
    \item $\Cov\big(z(\sbold_i),z(\sbold_j)\big)=\Cov\big(z(\sbold_i+\sbold_{\tau}),z(\sbold_j+\sbold_{\tau})\big).$
\end{itemize}
Under the stationary assumption the covariance function can be denoted by $C(\hbold),$ which emphasizes that the directed distance between the locations is the sufficient argument for the covariance function (rather than the coordinate values themselves). In many use cases, the raw data should be preprocessed in order to be stationary.

\begin{figure}[h!]
    \centering
        \includegraphics[width=1\linewidth]{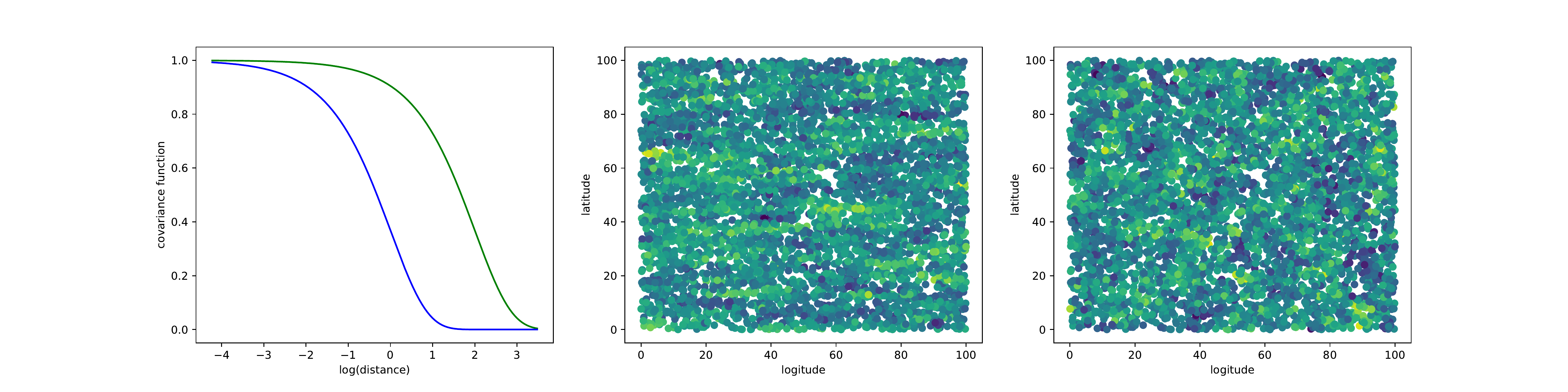}
         \caption{Simulated data of isotropy and anisotropy settings. Anisotropy setting: The left figure presents covariance functions in two directional axes that decay differently as a function of the distance (with $log_{10}$ scale). The green line is the covariance function for the longitudinal axis, and the blue line is for the latitudinal axis. The middle figure presents a simulated sample of an anisotropy setting. As once can see the variance in the latitudinal axis is larger than in the longitudinal axis. This is due to the higher correlation in the longitudinal axis than in the latitudinal axis. Isotropy setting: The right figure presents a simulated sample of an anisotropy setting.}\label{Figure: Demonstrating Settings}
    \end{figure}

A stronger assumption than stationarity is rotational invariance, $C(\hbold)=C(\|\hbold\|),$ which is called \textit{isotropy}. The isotropy assumption is taken frequently, also in cases it does not hold \citep{rajala2018review}, i.e., in \textit{anisotropic} settings. There are several anisotropy types, the most common is range anisotropy (also called geometric anisotropy), where $C(\hbold)$ decreases differently in different directions, but the sill, $\text{lim}_{\|\hbold\|\to\infty}C(\hbold),$ and the nugget effect, $\text{lim}_{\|\hbold\|\to 0}C(\hbold),$ do not depend on the direction. Figure \ref{Figure: Demonstrating Settings} demonstrates anisotropy and isotropy settings.

A specific type of range anisotropy is elliptic anisotropy, when a linear transformation of $\hbold$ induces isotropy. For example, the standard exponential kernel function, which assumes isotropy (see expression (\ref{Kernel: exponential, isotropy})), can be generalized using rotation and scaling matrices in order to capture elliptic anisotropy:
\begin{align}\label{Kernel: exponential, anisotropy}
   \sigma_{s}^2\times\exp{\big(-\|A\hbold\|}\big)+\sigma^2_{\epsilon}=\sigma_{s}^2\times\exp{\big(-\sqrt{\hbold^{t}A^{t}A\hbold}\big)}+\sigma^2_{\epsilon},   
\end{align}
where for $q=2:$
\[
A\coloneqq A(\lambda_1,\lambda_2,\eta)=
\begin{bmatrix} 
\frac{1}{\lambda_1} & 0 \\
0 & \frac{1}{\lambda_2} 
\end{bmatrix}
\begin{bmatrix} 
\cos(\eta) & -\sin(\eta) \\
\sin(\eta) & \cos(\eta) 
\end{bmatrix}
,
\]
$\eta\in[0,\pi]$  and $\eta+\pi/2$ are the anisotropy direction axes and $\lambda_i\in\mathbb{R}^{+}$ are the anisotropic scales. When $\lambda_1=\lambda_2$ expression (\ref{Kernel: exponential, anisotropy}) is reduced to expression (\ref{Kernel: exponential, isotropy}). 


Several hypothesis testing approaches for detecting anisotropy were proposed in recent years (for a literature review see Section \ref{section: Literature Review}), most of them focus on testing elliptic anisotropy. In many cases the tests are implemented on the variogram function which is a transformation of the covariance function:
\[
\gamma(\hbold_{i,j})=\frac{1}{2}\Var\big(z(\sbold_i)-z(\sbold_j)\big).
\]
It is easy to see that $\gamma(\hbold_{i,j})=C(0)-C(\hbold_{i,j}).$

Frequently, the variogram is estimated by the empirical variogram. The empirical variogram in the range $\hbold^*\pm\delta,$ where $\delta\in\mathbb{R}^d$ is:
\[
\widehat{\gamma}(\hbold)=\frac{1}{2\times|\mathcal{H}^*|}\sum_{(i,j)\in \mathcal{H}^*}\big(z(\sbold_i)-z(\sbold_j)\big)^2,\;\hbold\in\hbold^*\pm\delta\]
where $\mathcal{H}^*=\{(i,j)|\sbold_i-\sbold_j=\hbold^*\pm\delta\}$ and $|\mathcal{H}^*|$ is the size of the set $\mathcal{H}^*.$ 
In this paper we mainly refer to the covariance function.


\section{Our Approach}
 In this section we present our algorithms for testing anisotropy using re-sampling mechanisms. The first approach is based on parametric bootstrap hypothesis testing \citep{mackinnon2009bootstrap} and therefore requires parametric assumptions, the second approach is a non-parametric approach and therefore requires fewer assumptions.
\subsection{Parametric Bootstrap Based Test}\label{section: Parametric Bootstrap}
As in a standard hypothesis testing framework, two hypotheses are compared --- the null hypothesis ($H_0$) assumes that $\zbold(S)$ was sampled from a distribution with an isotropic covariance function, while the alternative hypothesis ($H_1$) assumes that $\zbold(S)$ was sampled from a distribution with an anisotropic covariance function. $H_1$ can specify the suspected anisotropic directional  axes, $\{\eta_i\}_{i\in[1,...,R]},$ or avoid specifying the anisotropic directional axes, but only posit that $R$ anisotropic directional axes exist. Another setting is when $H_1$ specifies ranges of $\{\eta_i\}_{i\in[1,...,R]}.$ For example, using the exponential kernel function family, elliptical anisotropy can be tested when $H_0$ assumes that the covariance function is the standard exponential kernel (expression (\ref{Kernel: exponential, isotropy})), while $H_1$ assumes that the covariance function is the elliptical exponential kernel function, presented in expression (\ref{Kernel: exponential, anisotropy}).
Other properties, such as stationarity and normality of $\zbold(S),$ are assumed equally by both hypotheses. The normality assumption will be relaxed in Section \ref{section: Non-Parametric Rotational Sampling Test}. 

Algorithm \ref{Algorithm: parametric} presents our parametric bootstrap hypothesis testing for anisotropy detection.
\begin{algorithm}[ht!]
\caption{Parametric Bootstrap Hypothesis Testing for Anisotropy Detection}
\label{Algorithm: parametric}
\begin{algorithmic}[1]
\STATE {\bfseries Input:} $\{S,\zbold(S)\}.$
\STATE {\bfseries Output:} P-value.
\STATE  Estimate $\mathcal{K}_{\thetabold|H_0}(\cdot),\;\mathcal{K}_{\thetabold|H_1}(\cdot)$ and denote their covariance matrices scores at $S$ as $\widehat{\Sigma}_{\thetabold|H_0},\;\widehat{\Sigma}_{\thetabold|H_1},$ respectively.
\STATE Calculate the following anisotropic discrepancy measure:
    \[\phi=\ell\big(\zbold;\widehat{\Sigma}_{\thetabold|H_1}\big)-\ell\big(\zbold;\widehat{\Sigma}_{\thetabold|H_0}\big),\]
    where $\ell\big(\zbold;\widehat{\Sigma}_{\thetabold|H_0}\big)$ and $\ell\big(\zbold;\widehat{\Sigma}_{\thetabold|H_1}\big)$ are the log-likelihood of $\zbold$ under the two hypotheses. 
\FOR{$b=1$ {\bfseries to} $B\in \mathbb{N}$}
 \STATE Sample one set of observations from $N_n(\mu\mathbbm{1},\widehat{\Sigma}_{\thetabold|H_0}),$ and denote the sample as $\zbold^{(b)}.$
\STATE Estimate $\mathcal{K}_{\thetabold|H_0}(\cdot),\;\mathcal{K}_{\thetabold|H_1}(\cdot)$ using $\zbold^{(b)}$ in the same way as in line 3 and denote their covariance matrices scores at $S$ as $\widehat{\Sigma}^{(b)}_{\thetabold|H_0},\;\widehat{\Sigma}^{(b)}_{\thetabold|H_1},$ respectively. 
\STATE Calculate 
    \[\phi^{(b)}=\ell\big(\zbold^{(b)};\widehat{\Sigma}^{(b)}_{\thetabold|H_1}\big)-\ell\big(\zbold^{(b)};\widehat{\Sigma}^{(b)}_{\thetabold|H_0}\big).\]
  \ENDFOR
\STATE  \[\text{P-value}=|\{\phi\leq\phi^{(b)}|b\in[1,...,B]\}|/B,\] where $|\cdot|$ is the set size.
\end{algorithmic}
\end{algorithm}

Technical details for Algorithm \ref{Algorithm: parametric}:
\begin{itemize}
    \item Line 3: The kernel parameters can be estimated using various approaches, such as maximum likelihood and restricted maximum likelihood (REML) of  $\zbold.$ More details about kernel parameters estimation can be found in \citet{rasmussen2003gaussian}.
    \item Line 4: $\phi,$ the anisotropic discrepancy measure, can also be written as follows:
  \[\phi=\Big(-\ell\big(\zbold;\widehat{\Sigma}_{\thetabold|H_0}\big)\Big)-\Big(-\ell\big(\widehat{\Sigma}_{\thetabold|H_1}\big)\Big).\] 
  Therefore $\phi$ is the loss function of the null hypothesis minus the loss function of the alternative hypothesis (where the loss function in this case is minus log-likelihood). 
  \item Line 5: $B$ controls the P-value resolution. For example, when $B=200$ the P-value resolution is $0.005.$ Also, $\Var(\text{P-value})$ decreases with $B$ (still, commonly the main factor affecting on $\Var(\text{P-value})$ is $\Var(\zbold(S)),$ rather than $B$).
  \item Line 6.: $\mu$ can be estimated by the mean. Alternatively, the data can be normalized, such that $\mu=0.$
  \item Line 10: By definition, the P-value is the probability of rejecting $H_0$ when $H_0$ is true, i.e., $P(\text{rejecting } H_0|H_0).$  Here, the distribution of $\phi$ under the null, $P(\phi|H_0),$ is numerically estimated (rather than derived analytically) and therefore the algorithm's output is $P(\text{rejecting } H_0|\widehat{H_0}).$ This is an inherent property in sampling-based hypothesis testing approaches.
  \end{itemize}

The logic behind Algorithm \ref{Algorithm: parametric} is that $P(\phi|H_0)$ is estimated using parametric bootstrap approach, i.e., by simulating $\{\phi^{(b)}\}_{b\in[1,...,B]}$ using the best empirical parametric representation of the data under the null hypothesis, $N_n(\mu\mathbbm{1},\widehat{\Sigma}_{\thetabold|H_0}).$ Then, $\phi,$ which is calculated using the data itself is compared to $\{\phi^{(b)}\}_{b\in[1,...,B]}$ in order to derive the P-value. Obviously, this framework involves significance parametric assumptions (normality and a specific kernel function family), however as will be presented in Section \ref{section: Literature Review}, these assumptions are relatively mild, and do not involve arbitrary hyperparameters and asymptotic approximations. Also, the parametric bootstrap hypothesis testing approach allows flexibility in different aspects. First, the statistic $\phi$ can be modified to other loss functions measuring the anisotropic discrepancy magnitude, such as test set error or even prediction errors, e.g., AIC \citep{akaike1974new}, Cp \citep{mallows1973some} and cross-validation \citep{stone1974cross} error types.
Secondly, Algorithm \ref{Algorithm: parametric} allows controlling $B,$ which tradeoffs between the resultant P-value resolution and $\Var(\text{P-value})$ on one hand and the computational cost on the other hand.

We note that our proposed approach is a straight forward application of parametric bootstrap to this problem, however to our knowledge this has not been previously proposed in the context of testing anisotropy. As we show below, it yields strong results even when the assumptions do not strictly apply. 

\subsubsection{Parametric Bootstrap with a Non-Parametric Covariance Function}\label{section: Parametric bootstrap with non-parametric covariance}
Besides normality, the main parametric assumption in Algorithm \ref{Algorithm: parametric} is the assumption of kernel function family. This section presents how this assumption can be potentially avoided by estimating the covariance function using a non-parametric regression with monotonicity constraints instead of a kernel function.

Since $y_{i,j}\coloneqq(z_i-\mu)\times (z_j-\mu)$ is an estimator of $C(\hbold_{i,j}),$ then $C(\hbold)$ can be estimated by modeling $\ybold=\{y_{i,j}\}_{(i,j)\in([1,...,n],[i,...,n])}$ (which contains $(n+1)\times n/2$ points) as a function of its corresponding location differences, $H=\{\hbold_{i,j}\}_{(i,j)\in([1,...,n],[i,...,n])}.$ The model should be constrained such that it yields non-negative values and the function is monotonically decreasing with respect to the distance (Tobler's law, \cite{tobler1970computer}). An example of a framework that facilitates enforcing monotonic constraints is isotonic regression \citep{stylianou2002dose,luss2014generalized}. Other popular tools that facilitate enforcing monotonic constraints are \href{https://stat.ethz.ch/R-manual/R-patched/library/mgcv/html/mono.con.html}{cubic regression splines model in mgcv package in R software} and \href{https://xgboost.readthedocs.io/en/latest/}{xgboost package in Python software}. Moreover, the idea of replacing the kernel function framework by monotonic regression in other applications is not new, for example see \cite{bjornstad2001nonparametric,wu2003nonparametric,choi2014modeling}.\footnote{Unfortunately, these studies do not supply code for implementation of their algorithm.}

Replacing the kernel function in Algorithm \ref{Algorithm: parametric} by a monotonic regression model requires specifying the suspected anisotropic directional axes $\{\eta_i\}_{i=1}^{R}.$ For example, when implementing isotonic regression in Algorithm \ref{Algorithm: parametric}, the monotonicity constraint in the null hypothesis model is forced on $\{\|\hbold_{i,j}\|\}_{(i,j)\in([1,...,n],[i,...,n])},$ while in the alternative hypothesis is forced on the projections of $H$ on  $\{\eta_i\}_{i=1}^{R}.$ 
Besides the covariance function estimation, all other parts in Algorithm \ref{Algorithm: parametric} remain the same.

We have invested substantial effort in implementing this approach with non-parametric covariance estimation, trying out the different tools mentioned above with no success. The main reason for this failure seems to be the difficulty of estimating the covariance matrices well. They are consistently poorly estimated and the resulting tests have very poor performance. 
Other issues which come up are:
\begin{itemize}
\item The fitted monotonic regression model might be a non strictly positive definite function, therefore the covariance matrix that is plugged-in the likelihood function can be non-positive definite (which make it as a illegitimate covariance matrix). A simple workaround for solving this issue is removing the  eigenvectors with the non-positive eigenvalues from the estimated covariance matrices.
\item As opposed to Section \ref{section: Parametric Bootstrap}, here the anisotropic directional axes must be prespecified. 
\item Models enforcing monotonicity are computationally expensive, especially models with multiple covariates, as in the anisotropic model. Taking into account the large sample size in our application, $(n+1)\times n/2,$ then when $n$ is not very small, the running time is long. Therefore, fitting the model using a sub-sample of $\{\ybold,H\}$ can be necessary sometimes. Note that since the elements in $\{\ybold,H\}$ are correlated, due to the overlap between the $z(\sbold_i)$'s elements  composing $\ybold$ and due to the spatial correlation between $z(\sbold_i)$'s themselves, then the effective sample size is smaller than the nominal one.
\end{itemize}

\subsection{Non-Parametric Rotational Sampling Test}\label{section: Non-Parametric Rotational Sampling Test}
Algorithm \ref{Algorithm: parametric} is based on the normality assumption, which is very common in spatial regression of a continuous dependent variable, however in many cases the deviation from normality is prominent and it is not possible to assume normality \citep{horvath2020testing}. 
Here we present a different approach which relies on the rotational symmetry that exists under the null hypothesis, to test it with no distributional assumptions. However, the suspected anisotropic directions have to be specified in advance (exactly or at least approximately). This approach is presented in Algorithm \ref{Algorithm: non-parametric}. For improved readability, we present Algorithm \ref{Algorithm: non-parametric}
where two perpendicular anisotropic directional axes are specified, however as will be demonstrated in Section \ref{section: Real Data}, it can be easily generalized for more than two and non-perpendicular anisotropic directional axes.
\begin{algorithm}[ht!]
\caption{Non-Parametric Rotational Sampling Hypothesis Testing for Anisotropy Detection}
\label{Algorithm: non-parametric}
\begin{algorithmic}[1]
\STATE {\bfseries Input:} $\{S,\zbold(S)\},\;\eta,\;\alpha\in[0,\pi/4).$
\STATE {\bfseries Output:} P-value. 
\STATE Derive $\{\ybold,H\}$ using $\{S,\zbold(S)\}.$
\STATE  Calculate
\begin{align*}
\phi_{isotropy}&=\underset{\thetabold|H_0}{\min} \sum_{i=1}^{n}\sum_{j=i}^{n}\big(y_{i,j}-\mathcal{K}_{\thetabold|H_0}(\|\hbold_{i,j}\|)\big)^2\\
\phi_{anisotropy}&=\underset{\thetabold|H_1}{\min} \sum_{i=1}^{n}\sum_{j=i}^{n}\big(y_{i,j}-\mathcal{K}_{\thetabold|H_1}(\hbold_{i,j})\big)^2
\end{align*}
\[\phi=\phi_{isotropy}-\phi_{anisotropy}.\]
\FOR{$b=1$ {\bfseries to} $B\in \mathbb{N}$}
 \STATE Sample random directional axis $\eta^{(b)}$ from  $[\eta+\alpha,\eta+\pi/2-\alpha].$
 \STATE Calculate
\[\phi^{(b)}_{anisotropy}=\underset{\thetabold|H_1,\eta^{(b)}}{\min} \sum_{i=1}^{n}\sum_{j=i}^{n}\big(y_{i,j}-\mathcal{K}_{\thetabold|H_1,\eta^{(b)}}(\hbold_{i,j})\big)^2,\]
where $\mathcal{K}_{\thetabold|H_1,\eta^{(b)}}$ is the anisotropic kernel with the anisotropic directional axes $\{\eta^{(b)},\eta^{(b)}+\pi/2\}$ instead of $\{\eta,\eta+\pi/2\}.$ 
\[\phi^{(b)}=\phi_{isotropy}-\phi^{(b)}_{anisotropy}.\]
  \ENDFOR
\STATE  \[\text{P-value}=|\{\phi\leq\phi^{(b)}|b\in[1,...,B]\}|/B.\]
\end{algorithmic}
\end{algorithm}


Technical details for Algorithm \ref{Algorithm: non-parametric}:
\begin{itemize}
    \item Line 4.: Since Algorithm \ref{Algorithm: non-parametric} enables avoiding assuming normality, it is more reasonable to use squared errors loss function than a likelihood-based loss function.
    \item Line 6.: In order to increase power, $\alpha$ prevents sampling axes that are close to the specified anisotropic directional axes.
    When the specified anisotropic directional axes are not perpendicular or the number of axes is larger than two, then the space of $\alpha$ is changed accordingly.
    \item Line 7.: Unlike in Algorithm \ref{Algorithm: parametric}, since a new data is not simulated, $\phi_{isotropy}$ can be reused in $\phi^{(b)}.$
\end{itemize}

As in Algorithm \ref{Algorithm: parametric}, Algorithm \ref{Algorithm: non-parametric}
also estimates $P(\phi|H_0)$ using simulation, however here the \textit{directional axes are simulated}, rather than parametric bootstrap samples. In that way the original data can also be used for calculating $\{\phi^{(b)}\}_{b\in[1,...,B]}$ and 
distribution assumptions can be avoided. On the other hand, sampling the directional axes forces specifying the suspected anisotropic directional axes (unlike in Algorithm \ref{Algorithm: parametric}).

The requirement for specifying suspected anisotropic directional axes can be partially relaxed, such that ranges that include the suspected anisotropic directional axes are specified instead of the exact values. It can be done by utilizing $\alpha$ for constructing non-overlapping domains of $\{\eta_i\}_{i\in[1,...,R]}$ and $\{\eta^{(b)}\}_{b\in[1,...,B]}.$ For example, in case the suspected anisotropic directional ranges are $\{\eta_1\in[-\alpha,\alpha],\eta_2=\eta_1+\pi/2\},$ then the estimation of $\phi_{anisotropy}$ should also include optimization of $\eta_1.$ Correspondingly, the sampling space in line 6 is $[2\times\alpha,\pi/2-2\times\alpha],$ and the optimization in line 7 is also over $\eta^{(b)*}\in[\eta^{(b)}-\alpha,\eta^{(b)}+\alpha].$ 

Similarly to Algorithm \ref{Algorithm: parametric}, the kernel can be potentially replaced by non-parametric monotonic regression. In that way, both main parametric assumptions --- normality and kernel structure --- are avoided.

\section{Literature Review}\label{section: Literature Review}
The most common approach for detecting anisotropy is by analyzing popular graphs, such as directional variograms plot and rose diagram \citep{weller2016review}, which compare empirical variograms with respect to different directional exes. Although graphs can be very informative, analyzing them is open to subjective interpretations and therefore cannot be used for deriving objective scientific conclusions. Another drawback in using these graphs, which also appears in many hypothesis testings methods for detecting anisotropy, is their reliance on hyperparameters (e.g., $\delta,$ see Section \ref{section: introduction}) which may affect the conclusion.

A popular hypothesis testing approach for detecting anisotropy is utilizing the asymptotic distribution of the empirical variogram (which was derived by \citet{baczkowski1987approximate}) in order to calculate the P-value of tests whose statistic is based on the empirical variograms. For example, in \cite{guan2004nonparametric}, lags, $[\hbold_1,....,\hbold_K],$ that relate to the suspected anisotropic directional axes are selected, such that under $H_0$ (which assumes isotropy) $AG=0,$ where $G=[\gamma(\hbold_1),....,\gamma(\hbold_K)]$ and A is the contrast matrix, while under $H_1,$ $AG\neq0.$ The statistic of the test is $c(A\widehat{G})^{t}(A\Sigma_GA^{t})^{-1}A\widehat{G},$ where $\widehat{G}=[\widehat{\gamma}(\hbold_1),....,\widehat{\gamma}(\hbold_K)],$ $\Sigma_G$ is the variance of the asymptotic distribution of $A\widehat{G}$ under $H_0$ and $c\in \mathbb{R}.$ The P-value can be derived by the asymptotic distribution of the statistic (which is  $\chi^2$).
The main drawbacks in this method is deriving $\Sigma_G$ and the selection of the smoothing parameter $\delta$ when calculating $\widehat{G}.$ Also, the asymptotic derivations in \cite{guan2004nonparametric} assume specific sampling mechanisms of $S,$ and therefore the test is not valid in many applications which do not follow their sampling mechanisms. \citet{maity2012testing} modify \citeauthor{guan2004nonparametric}'s method for testing isotropy in the covariance function framework by estimating the covariance function at $[\hbold_1,...,\hbold_K]$ using kernel regression. Also, they suggest to estimate the variance matrix of the asymptotic distribution using block bootstrap. Both methods, \cite{guan2004nonparametric} and \cite{maity2012testing}, do not rely on specification of a kernel covariance/variogram function, however their tests rely on selection of hyperparameters --- $\delta$ in \cite{guan2004nonparametric} and the kernel regression function and its bandwidth in \cite{maity2012testing}. \cite{weller2017nonparametric} claims that the resulting tests are very sensitive to the hyperparameters selection.



Detecting anisotorpy correlation structure in the spectral domain, where the  asymptotic derivations are simpler than in the spatial domain, is also common \citep{weller2016review}.
\cite{van2020general} propose a unified framework for detecting different properties of the covariance function, including anisotropy. Similarly to \cite{guan2004nonparametric} and \cite{maity2012testing}, \citeauthor{van2020general}'s test requires pre-specification of the directional axes, however the statistic assessing the deviation from isotropy is based on the periodogram, which is the corresponding object to variogram/covariance function in the spectral domain. Still, some challenges are common with \citeauthor{maity2012testing}'s approach. First, \citeauthor{van2020general} provide an asymptotic result, and it is unclear how their method performs for small sample sizes (the smallest sample size that is used in their simulations is $n=1200$). Also, this method requires selecting a set of lags and number and spacing of frequencies for calculating the statistic. 
Besides \cite{van2020general}, there are other papers that analyze covariance function properties, however many of them focus on different types of symmetry \citep{weller2017nonparametric}, which is closely related to isotropy (isotropy does imply symmetry, but symmetry does not imply isotropy).

\section{Simulation Experiments}\label{section: simulation}
This section presents simulation experiments which compare between Algorithm \ref{Algorithm: parametric}, Algorithm \ref{Algorithm: non-parametric} and the algorithm presented by \citet{maity2012testing}, which we refer to as MS.

\paragraph{Distributional setting}
The sample, $\{z(\sbold_i)\}_{i=1}^{n},$ was drawn from a normal distribution with zero mean and the elliptic  exponential kernel covariance function (see expression \ref{Kernel: exponential, anisotropy}) with the parameters $\sigma_{\epsilon}^2=1,\sigma_s^2=1,\eta=0,\lambda_1=1$ and different $\lambda_2$ values. The coordinate values, $\{\sbold_i\}_{i=1}^{n},$ were drawn from the uniform distribution $U(0,1).$ The simulation experiment was executed for different sample sizes, $n=200/500/1000,$ and different anisotropic intensity, $\lambda_2=1/2/5/10$ ($\lambda_2=1$ means an isotropic kernel). Also, the number of resamples $B$ was set to $200.$ For the simulation code see at \textbf{(anonymized for review)}.

\paragraph{Hypotheses definition} In all the three algorithms the null hypothesis assumes isotropy (see expression \ref{Kernel: exponential, isotropy}) and the alternative hypothesis specifies the anisotropic directional axes $\{\eta_1=0,\eta_2=\pi/2\}.$
As presented in Section \ref{section: Parametric Bootstrap}, Algorithm \ref{Algorithm: parametric} enables avoiding specifying anisotropic directional axes, however in order to present a fair comparison between all the algorithms, the hypotheses are defined here in the same way for all the algorithms. For other settings, where the anisotropic directional axes are not specified for Algorithm \ref{Algorithm: parametric}, see Section \ref{section: Real Data}.

\paragraph{Technical details}
For Algorithm \ref{Algorithm: parametric}, the \href{https://scikit-learn.org/stable/modules/gaussian_process.html}{Gaussian Process module} from the sklearn package in python was used to estimate the kernel parameters. For Algorithm \ref{Algorithm: non-parametric}, in order to reduce running time, the test was implemented only on $10,000$ observations which were drawn without replacement from $\{\ybold,H\}.$ Also, $\alpha$ was set to $\pi/36$ (which is $5^o$). For MS, SpTest package in R software  \citep{weller2015sptest} was used with the parameters:
$\text{nBoot}=75,\;\text{blockdims}=\{1,1\},\;\text{grid}=\{0.1,0.1\},$ defaulted lags array of $\Lambda=\{\hbold_1=(0.1,0),\;\hbold_2=(0,0.1),\;\hbold_1=(0.1,0.1),\;\hbold_1=(-0.1,0.1)\}$
and the contrast matrix $A=\begin{bmatrix}
1 & -1 & 0&0\\
0 & 0 & 1&-1
\end{bmatrix}.$ Each simulation setting was repeated about $200$ times in order to estimate the P-value distribution.

The following table presents the estimated Type I error and power under significance level of $0.05,$ (i.e., $\widehat{P}(\text{P-value}<0.05)$) for the different settings.

\begin{table}\footnotesize\label{Table: power}
\caption{The table presents empirical power for different settings for significance level of $0.05.$ The $\lambda_2=1$ column is the Type I error estimates.}
\fbox{%
\begin{tabular}{|l|ccc|ccc|ccc|ccc|}
                           & \multicolumn{3}{c|}{$\lambda_2=1$}                              & \multicolumn{3}{c|}{$\lambda_2=2$}                           & \multicolumn{3}{c|}{$\lambda_2=5$}                            & \multicolumn{3}{c|}{$\lambda_2=10$}       \\
Algorithm\textbackslash{}n & 200                        & 500                        & 1000  & 200                       & 500                       & 1000 & 200                       & 500                       & 1000  & 200                       & 500                       & 1000  \\
Algorithm \ref{Algorithm: parametric}                         & 0.05 & 0.04 & 0.05 & 0.11	&0.31	&0.43
& 0.47	&0.70	&0.94&0.65&0.89&	0.99 \\ 
Algorithm \ref{Algorithm: non-parametric}                         &0.14&	0.07&	0.07 & 0.05&0.15	&0.19& 0.16&	0.29&	0.24
& 0.27&	0.30&	0.33
  \\ 
MS  &0.02&	0.04&	0 & 0.06	&0.04 &0.05 & 0.02	&0.10 &0.10&  0.07&	0.10&	0.17  \\
\end{tabular}}
\end{table}



As we can see, Algorithm \ref{Algorithm: parametric} has the highest power. It is not surprising, since Algorithm \ref{Algorithm: parametric} makes a stronger distributional assumption --- normality of $\zbold$ --- which holds in our setting. Also, as expected, the power of all the three algorithms increases with the sample size and $\lambda_2,$ however the power of MS is relatively low in our settings. Also, as we can see in the $\lambda_2=1$ column, Algorithm \ref{Algorithm: parametric} controls the Type I error properly for significance level of $0.05,$ Algorithm \ref{Algorithm: non-parametric} also controls the Type I error properly for moderate and large sample (but not for $n=200$) and MS does not control Type I error properly.

\begin{figure}[h]
    \centering
        \includegraphics[width=0.8\linewidth]{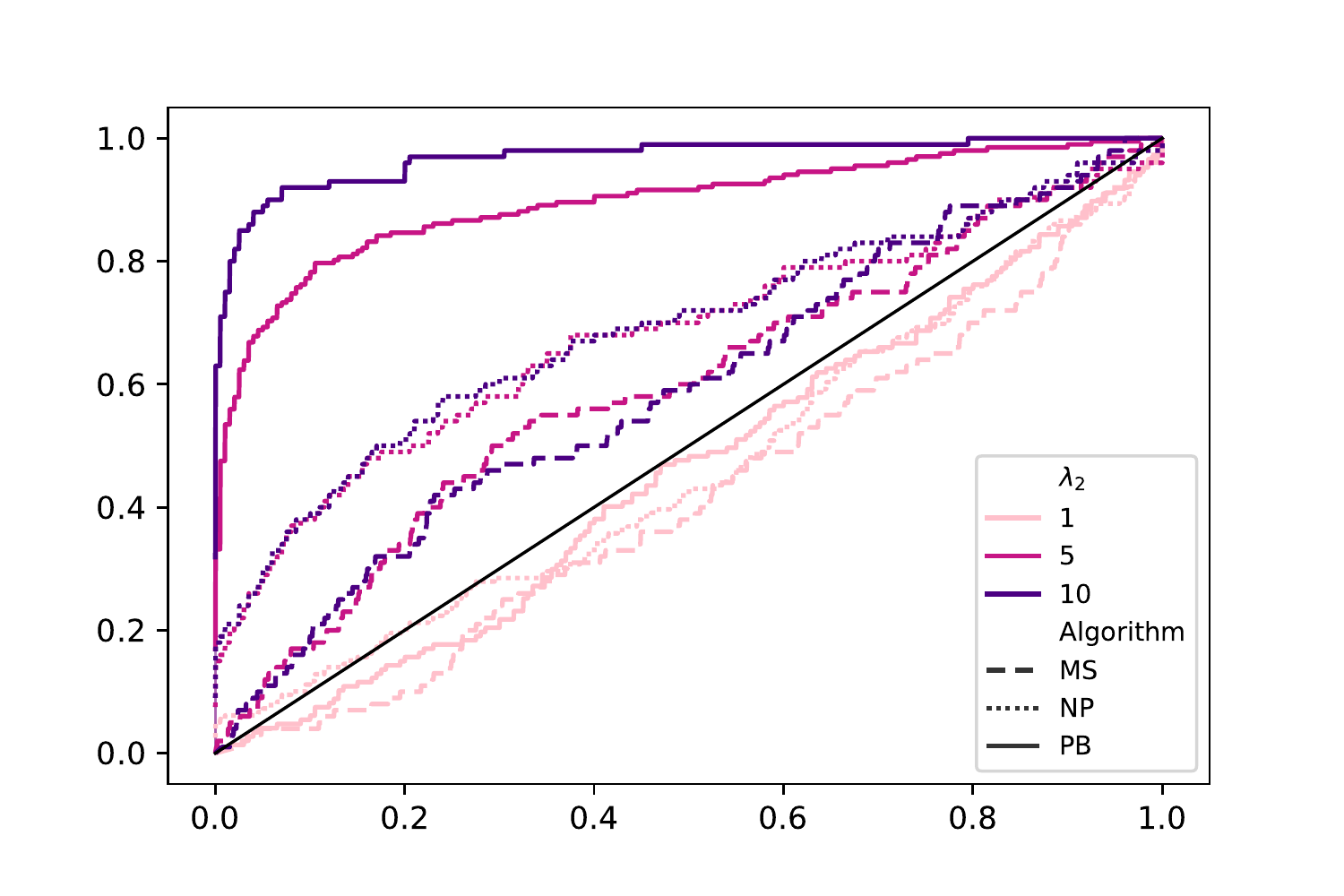}
         \caption{P-value empirical cumulative distribution for Algorithms \ref{Algorithm: parametric}, \ref{Algorithm: non-parametric} and MS for $n=500$ and different  $\lambda_2$ values.}\label{Figure: power for n=500}
    \end{figure}

Figure \ref{Figure: power for n=500} presents the empirical cumulative distribution of the P-value of the three algorithms for $n=500$ and for $\lambda_2=1/5/10.$
    
As we can see, Algorithms \ref{Algorithm: parametric} has a higher power than the other algorithms for any significance level. Still, Algorithms \ref{Algorithm: non-parametric} also has a much higher power than MS. 

\section{Real Data Analysis}\label{section: Real Data}
This section analyzes and demonstrates implementation of algorithms \ref{Algorithm: parametric}, \ref{Algorithm: non-parametric} and MS in various use cases using four datasets that were introduced by \cite{hohn1998geostatistics}. The datasets can be found in the \href{http://www.wvgs.wvnet.edu/www/geostat/GeostatPetGeol.html}{book's website}. The code can be found at \textbf{(anonymized for review)}.

\begin{figure}[h]
    \centering
        \includegraphics[width=1\linewidth]{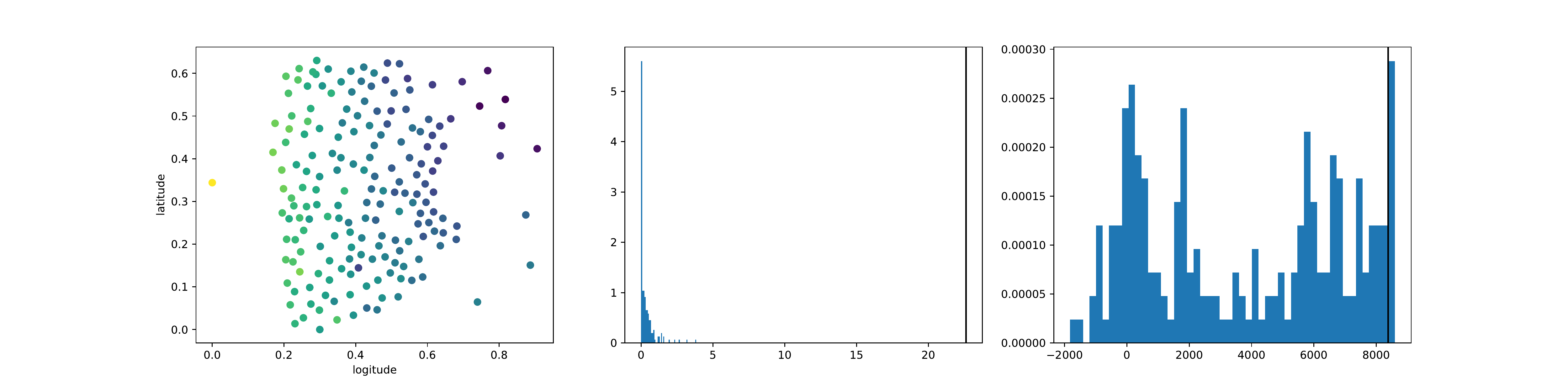}
         \caption{Granny Creek Field. The left figure is the scatter plot of the Granny Creek Field dataset after scaling, where the color intensity indicates the elevation value. The middle and the right figures compare the $\phi$ value (in black) with $\{\phi^{(b)}\}_{b=1}^{200}$ for Algorithm \ref{Algorithm: parametric} and Algorithm \ref{Algorithm: non-parametric}, respectively.}\label{Figure: Granny Creek Field}
    \end{figure}

\paragraph{Granny Creek Field Data} 
The Granny Creek Field dataset contains $181$ measurements of a sandstone base elevation in Granny Creek Field, central West Virginia.
Figure \ref{Figure: Granny Creek Field} presents the analyzed standard score elevation measurements, $\zbold(S),$ and their standardized coordinates, $S.$ Based on prior geographical knowledge, \citeauthor{hohn1998geostatistics} claims that the expected anisotropic directional axes are $\{0,\pi/2\}.$ After analyzing the directional variograms graph, \citeauthor{hohn1998geostatistics} suggests that the anisotropic directional axes are $\{\eta_1=10\times\pi/18,\eta_2=\pi/9\}$ (10 and 100 degrees). Here we use Algorithms \ref{Algorithm: parametric}, \ref{Algorithm: non-parametric} and MS for testing whether $\{\eta_1=0,\eta_2=\pi/2\}$ are anisotropic directional axes. Using exponential kernel and setting $B=200,$ the P-value of Algorithm \ref{Algorithm: parametric} is smaller than $0.005$ and the P-value of Algorithm \ref{Algorithm: non-parametric} with $\alpha=\pi/36$ is $0.025.$ Figure \ref{Figure: Granny Creek Field} compares the $\phi$ value with $\{\phi^{(b)}\}_{b=1}^{200}$ for both algorithms. Implementing MS algorithm using the same setting as in Section \ref{section: simulation}, yields $\text{P-value}=0.011.$ Thus, all three testing approaches support anisotropy. 

\begin{figure}[h]
    \centering
        \includegraphics[width=1\linewidth]{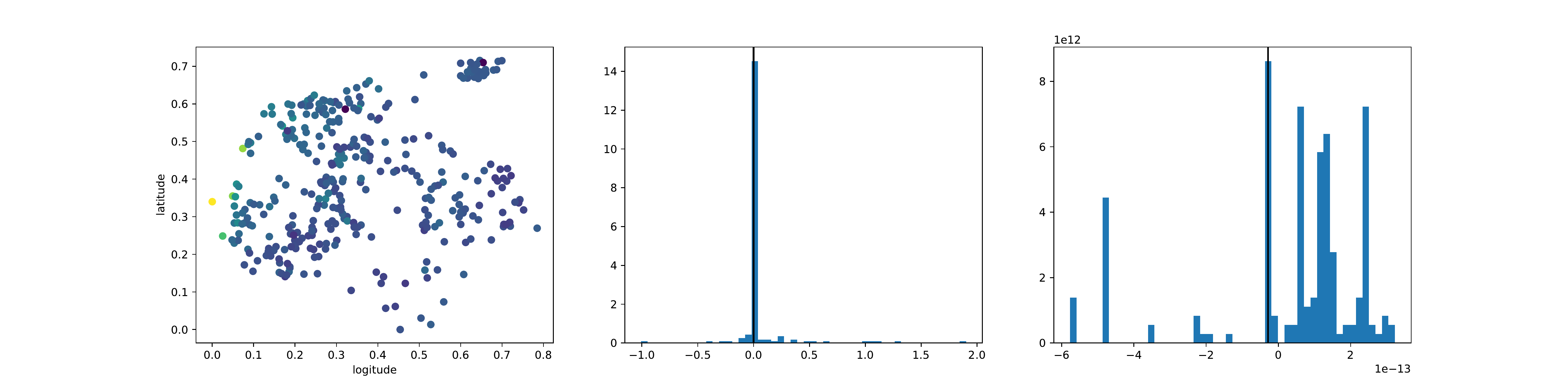}
         \caption{Mississippian Sandstone. The left figure is the scatter plot of the Mississippian Sandstone dataset after scaling, where the color intensity indicates the depth value. The middle and the right figures compare the $\phi$ value (in black) with $\{\phi^{(b)}\}_{b=1}^{200}$ for Algorithm \ref{Algorithm: parametric} and Algorithm \ref{Algorithm: non-parametric}, respectively.}\label{Figure: Mississippian Sandstone}
    \end{figure}

\paragraph{Mississippian Sandstone Data} 
The Mississippian Sandstone dataset contains
$348$ measurements of subsea depth of a mississippian-age reservoir sandstone base in Ritchie County, West Virginia. Besides standard scaling, an outlier observation was removed from the analysis. The data is presented in Figure \ref{Figure: Mississippian Sandstone}).

Based on prior geographical knowledge, \citeauthor{hohn1998geostatistics} claims that the correlation in the $\{\pi/4,\;3\times\pi/4\}$ directional axes is suspected to be different than the correlation in the $\{\pi,\;\pi/2\}$ directional axes. Therefore, the elliptical transformation of the anisotropic kernel (see equation \ref{Kernel: exponential, anisotropy}) is:
\[
A=
\begin{bmatrix} 
\frac{1}{\lambda_1} & 0& 0& 0 \\
0 & \frac{1}{\lambda_1} & 0& 0\\
0 & 0 &\frac{1}{\lambda_2}& 0\\
0 & 0 &0&\frac{1}{\lambda_2}\\
\end{bmatrix}
\begin{bmatrix} 
\cos(0) & \sin(0) \\
\cos(\frac{\pi}{2}) & \cos(\frac{\pi}{2}) \\
\cos(\frac{\pi}{4}) & \sin(\frac{\pi}{4}) \\
\cos(\frac{3\pi}{4}) & \cos(\frac{3\pi}{4}) \\
\end{bmatrix},
\]
i.e., $\lambda_1$ is the length-scale parameter of $\{\eta_3=0,\eta_4=\pi/2\}$ and $\lambda_2$ is the length-scale parameter of $\{\eta_1=\pi/4,\eta_2=3\times\pi/4\}.$ 

The P-value for Algorithm \ref{Algorithm: parametric} is $0.93$ and for Algorithm \ref{Algorithm: non-parametric}  with $\alpha=\pi/72$ is $0.7.$ 
Figure \ref{Figure: Mississippian Sandstone} compares the $\phi$ value with $\{\phi^{(b)}\}_{b=1}^{200}.$ Implementing MS algorithm with 
contrast matrix, $A=\begin{bmatrix}
1 & 0 &-1&0\\
0 & 1 & 0&-1\\
1 & 1 & -1&-1\\
\end{bmatrix},$ (other parameters are the same as in Section \ref{section: simulation}), gives $\text{P-value}=0.19.$

\begin{figure}[h]
    \centering
        \includegraphics[width=1\linewidth]{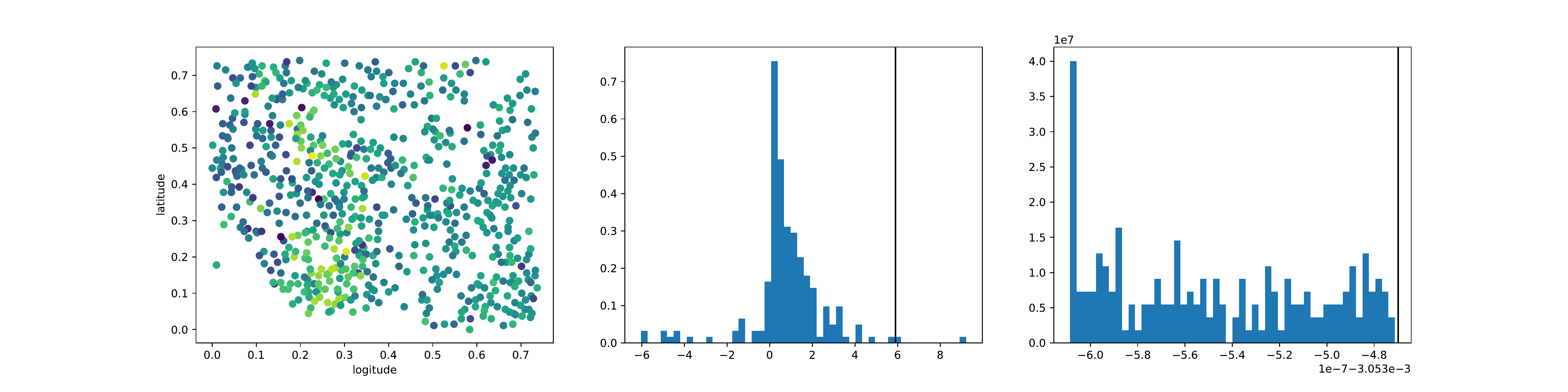}
         \caption{Gas potential in Barbour County. The left figure is the scatter plot of the gas potential in Barbour County dataset after scaling, where the color intensity indicates the gas potential. The middle and the right figures compare the $\phi$ value (in black) with $\{\phi^{(b)}\}_{b=1}^{200}$ for Algorithm \ref{Algorithm: parametric} and Algorithm \ref{Algorithm: non-parametric}, respectively.}\label{Figure: Barbour County Data}
    \end{figure}

\paragraph{Gas Potential in Barbour County Data} This data contains  $674$ measurements of gas initial potential, producing from upper Devonian sandstones and siltstones in Barbour County, West Virginia. Based on \citeauthor{hohn1998geostatistics}'s suggestion, in order to satisfy the normality assumption (in Algorithm \ref{Algorithm: parametric}), the standard score of the log of the gas initial potential is analyzed. Figure \ref{Figure: Barbour County Data} presents the data before and after pre-processing. In this example there is no prior information about suspected anisotropic directional axes. Implementing Algorithm \ref{Algorithm: parametric} without specifying suspected anisotropic directional axes (however assuming that there are two anisotropic directional axes) gives $\text{P-value}=0.015.$ Implementing Algorithms \ref{Algorithm: non-parametric} and specifying defaulted anisotropic directional axes of $\{\eta_1=0,\eta_2=\pi/2\}$ with $\alpha=\pi/36$ gives $\text{P-value}<0.005.$ A comparison of the $\phi$ value with $\{\phi^{(b)}\}_{b\in[1,...,200]}$ is presented in Figure \ref{Figure: Barbour County Data}. Implementing MS algorithm using the same setting given in Section \ref{section: simulation}, gives $\text{P-value}=0.01.$



\begin{figure}[h]
    \centering
        \includegraphics[width=1\linewidth]{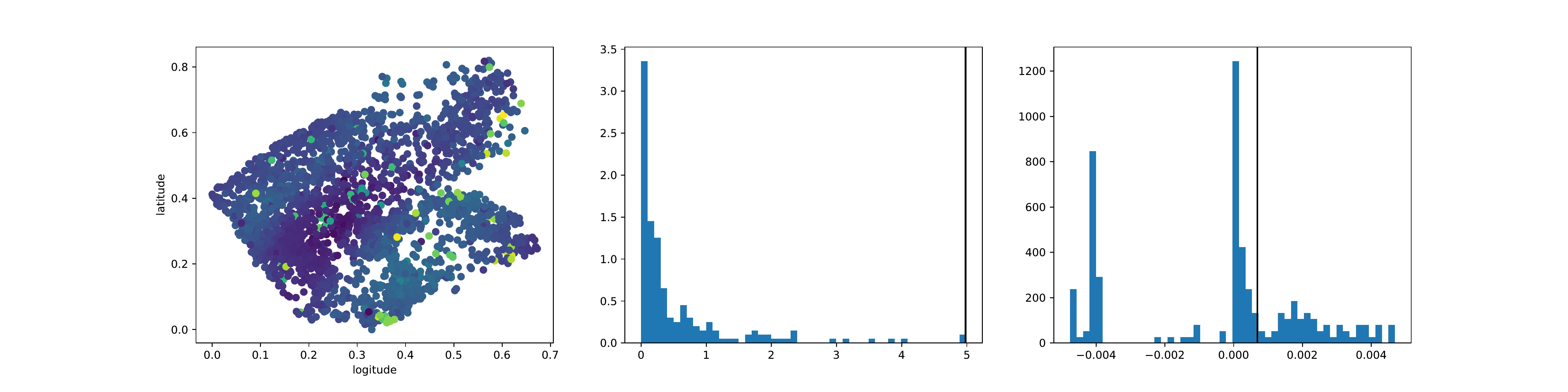}
         \caption{Greenbrier limestone. The left figure is the scatter plot of the greenbrier limestone dataset after scaling, where the color intensity indicates the elevation values. The middle and the right figures compare the $\phi$ value (in black) with $\{\phi^{(b)}\}_{b=1}^{200}$ for Algorithm \ref{Algorithm: parametric} and Algorithm \ref{Algorithm: non-parametric}, respectively.}\label{Figure: Greenbrier Limestone}
    \end{figure}

\paragraph{Greenbrier Limestone}
The Greenbrier Limestone dataset contains measurements at $2,335$ tops of the Greenbrier
Limestone. Based on graphical analysis, 
\citeauthor{hohn1998geostatistics} claims that the anisotropic directional axes are \{$\eta_1=\pi/3,\;\eta_2=2\times\pi/3$\}. Since using graphical analyses for selecting the suspected anisotorpic directional axes of the alternative hypothesis violates our hypothesis testing scheme (as well as many other standard hypothesis testing schemes), here we use the default anisotropic directional axes, $\{\eta_1=0,\eta_2=\pi/2\}.$ Also, unlike in the analysis of the Gas Potential in Barbour Country dataset, these axes will also be specified for Algorithms \ref{Algorithm: parametric}. The P-value of Algorithm \ref{Algorithm: parametric} is $0.005$ and the P-value of Algorithm \ref{Algorithm: non-parametric} with $\alpha=\pi/36$ is $0.3.$ Figure \ref{Figure: Greenbrier Limestone} compares the $\phi$ value with $\{\phi^{(b)}\}_{b\in[1,...,B]}.$ Implementing MS algorithm using the same setting given in Section \ref{section: simulation} but with $grid=[0.5,0.5]$ (due to convergence issue), gives $\text{P-value}=0.12.$ Therefore, as we can see, while Algorithm \ref{Algorithm: parametric} rejects the null hypothesis with significance level of $0.05,$ Algorithm \ref{Algorithm: parametric} and MS do not reject.

\section{Summary and Conclusions}
In this paper we introduce and analyze parametric and non-parametric hypothesis testing algorithms for detecting anisotropic covariance structure. In both algorithms, the statistic distribution under the null hypothesis is obtained using resampling mechanisms, while asymptotic parametric approximations and hyperparameters selection, which are the Achilles' heel of other methods, are avoided.

Both algorithms propose general frameworks that can also be applied when the covariance function is estimated using a non-parametric monotonic regression model, which enables avoiding assuming a parametric kernel function. However, as explained in Section \ref{section: Parametric bootstrap with non-parametric covariance}, modeling the covariance function using a non-parametric monotonic regression model is a difficult task and requires further research. 

The numerical results demonstrate the high power that is obtained by the proposed algorithms, and also demonstrates how the algorithms can easily be adjusted in various settings. In our view, the simple resampling-based methods we propose here should be considered as leading practical approaches for testing anisotropy.

\bibliography{Reference}

\begin{thebibliography}{24}
\expandafter\ifx\csname natexlab\endcsname\relax\def\natexlab#1{#1}\fi
\expandafter\ifx\csname url\endcsname\relax
  \def\url#1{\texttt{#1}}\fi
\expandafter\ifx\csname urlprefix\endcsname\relax\def\urlprefix{URL: }\fi

\bibitem[{Akaike(1974)}]{akaike1974new}
Akaike, H. (1974) A new look at the statistical model identification.
\newblock \textit{IEEE Transactions on Automatic Control}, \textbf{19},
  716--723.

\bibitem[{Baczkowski and Mardia(1987)}]{baczkowski1987approximate}
Baczkowski, A. and Mardia, K. (1987) Approximate lognormality of the sample
  semi-variogram under a gaussian process.
\newblock \textit{Communications in Statistics-Simulation and Computation},
  \textbf{16}, 571--585.

\bibitem[{Bj{\O}rnstad and Falck(2001)}]{bjornstad2001nonparametric}
Bj{\O}rnstad, O.~N. and Falck, W. (2001) Nonparametric spatial covariance
  functions: estimation and testing.
\newblock \textit{Environmental and Ecological Statistics}, \textbf{8}, 53--70.

\bibitem[{Choi(2014)}]{choi2014modeling}
Choi, I. (2014) \textit{Modeling spatial covariance functions}.
\newblock Ph.D. thesis, Purdue University.

\bibitem[{Guan(2004)}]{guan2004nonparametric}
Guan, Y.~T. (2004) \textit{Nonparametric methods of assessing spatial
  isotropy}.
\newblock Ph.D. thesis, Texas A\&M University.

\bibitem[{Hohn(1998)}]{hohn1998geostatistics}
Hohn, M. (1998) \textit{Geostatistics and petroleum geology}.
\newblock Springer Science \& Business Media.

\bibitem[{Horv{\'a}th et~al.(2020)Horv{\'a}th, Kokoszka and
  Wang}]{horvath2020testing}
Horv{\'a}th, L., Kokoszka, P. and Wang, S. (2020) Testing normality of data on
  a multivariate grid.
\newblock \textit{Journal of Multivariate Analysis}, \textbf{179}, 104640.

\bibitem[{Li and Chen(2016)}]{li2016review}
Li, P. and Chen, S. (2016) A review on gaussian process latent variable models.
\newblock \textit{CAAI Transactions on Intelligence Technology}, \textbf{1},
  366--376.

\bibitem[{Luss and Rosset(2014)}]{luss2014generalized}
Luss, R. and Rosset, S. (2014) Generalized isotonic regression.
\newblock \textit{Journal of Computational and Graphical Statistics},
  \textbf{23}, 192--210.

\bibitem[{MacKinnon(2009)}]{mackinnon2009bootstrap}
MacKinnon, J.~G. (2009) Bootstrap hypothesis testing.
\newblock \textit{Handbook of computational econometrics}, \textbf{183}, 213.

\bibitem[{Maity and Sherman(2012)}]{maity2012testing}
Maity, A. and Sherman, M. (2012) Testing for spatial isotropy under general
  designs.
\newblock \textit{Journal of statistical planning and inference}, \textbf{142},
  1081--1091.

\bibitem[{Mallows(1973)}]{mallows1973some}
Mallows, C.~L. (1973) Some comments on cp.
\newblock \textit{Technometrics}, \textbf{15}, 661--675.

\bibitem[{Mihoub et~al.(2016)Mihoub, Chabour and Guermoui}]{mihoub2016modeling}
Mihoub, R., Chabour, N. and Guermoui, M. (2016) Modeling soil temperature based
  on gaussian process regression in a semi-arid-climate, case study ghardaia,
  algeria.
\newblock \textit{Geomechanics and Geophysics for Geo-Energy and
  Geo-Resources}, \textbf{2}, 397--403.

\bibitem[{Rajala et~al.(2018)Rajala, Redenbach, S{\"a}rkk{\"a} and
  Sormani}]{rajala2018review}
Rajala, T., Redenbach, C., S{\"a}rkk{\"a}, A. and Sormani, M. (2018) A review
  on anisotropy analysis of spatial point patterns.
\newblock \textit{Spatial Statistics}, \textbf{28}, 141--168.

\bibitem[{Rasmussen(2003)}]{rasmussen2003gaussian}
Rasmussen, C.~E. (2003) Gaussian processes in machine learning.
\newblock In \textit{Summer school on machine learning}, 63--71. Springer.

\bibitem[{Stone(1974)}]{stone1974cross}
Stone, M. (1974) Cross-validatory choice and assessment of statistical
  predictions.
\newblock \textit{Journal of the Royal Statistical Society: Series B
  (Methodological)}, \textbf{36}, 111--133.

\bibitem[{Stylianou and Flournoy(2002)}]{stylianou2002dose}
Stylianou, M. and Flournoy, N. (2002) Dose finding using the biased coin
  up-and-down design and isotonic regression.
\newblock \textit{Biometrics}, \textbf{58}, 171--177.

\bibitem[{Tobler(1970)}]{tobler1970computer}
Tobler, W.~R. (1970) A computer movie simulating urban growth in the detroit
  region.
\newblock \textit{Economic geography}, \textbf{46}, 234--240.

\bibitem[{Van~Hala et~al.(2020)Van~Hala, Bandyopadhyay, Lahiri, Nordman
  et~al.}]{van2020general}
Van~Hala, M., Bandyopadhyay, S., Lahiri, S.~N., Nordman, D.~J. et~al. (2020) A
  general frequency domain method for assessing spatial covariance structures.
\newblock \textit{Bernoulli}, \textbf{26}, 2463--2487.

\bibitem[{Weller(2015)}]{weller2015sptest}
Weller, Z.~D. (2015) Sptest: An r package implementing nonparametric tests of
  isotropy.
\newblock \textit{arXiv preprint arXiv:1509.07185}.

\bibitem[{Weller(2017)}]{weller2017nonparametric}
--- (2017) \textit{Nonparametric tests of spatial isotropy and a
  calibration-capture-recapture model}.
\newblock Ph.D. thesis, Colorado State University. Libraries.

\bibitem[{Weller et~al.(2016)Weller, Hoeting et~al.}]{weller2016review}
Weller, Z.~D., Hoeting, J.~A. et~al. (2016) A review of nonparametric
  hypothesis tests of isotropy properties in spatial data.
\newblock \textit{Statistical Science}, \textbf{31}, 305--324.

\bibitem[{Wu and Pourahmadi(2003)}]{wu2003nonparametric}
Wu, W.~B. and Pourahmadi, M. (2003) Nonparametric estimation of large
  covariance matrices of longitudinal data.
\newblock \textit{Biometrika}, \textbf{90}, 831--844.

\bibitem[{Zhong et~al.(2008)Zhong, Lotte, Girolami and
  L{\'e}cuyer}]{zhong2008classifying}
Zhong, M., Lotte, F., Girolami, M. and L{\'e}cuyer, A. (2008) Classifying eeg
  for brain computer interfaces using gaussian processes.
\newblock \textit{Pattern Recognition Letters}, \textbf{29}, 354--359.

\end{thebibliography}
\bibliographystyle{rss}

\end{document}